\documentclass[pra,showpacs,showkeys,twocolumn]{revtex4-1}

\usepackage{amsmath}
\usepackage{amsfonts}            
\usepackage{amssymb}

\usepackage{mathrsfs}

\usepackage{eufrak}

\usepackage{graphicx}
\usepackage{epstopdf}

\usepackage{dcolumn}
\usepackage{bm}

\makeatletter
\def\@dotsep{4.5}
\makeatother

\begin{document}


\title{Multiple scattering of matter waves: \\ an analytic model of the refractive index for atomic and molecular gases}

\author{Mikhail Lemeshko}
\email{mikhail.lemeshko@gmail.com}

\author{Bretislav Friedrich}

\email{brich@fhi-berlin.mpg.de}

\affiliation{%
Fritz-Haber-Institut der Max-Planck-Gesellschaft, Faradayweg 4-6, D-14195 Berlin, Germany
}%

\date{\today}

\begin{abstract}

We present an analytic model of the refractive index for matter waves propagating through atomic or molecular gases. The model, which combines a WKB treatment of the long range attraction with the Fraunhofer model treatment of the short range repulsion, furnishes a refractive index in compelling agreement with recent experiments of Jacquey \textit{et al.} [Phys. Rev. Lett. \textbf{98}, 240405 (2007)] on Li atom matter waves passing through dilute noble gases. We show that the diffractive contribution, which arises from scattering by a two dimensional ``hard core'' of the potential, is essential for obtaining a correct imaginary part of the refractive index.

\end{abstract}


\maketitle

\section{Introduction}

Multiple scattering of matter waves arises in a broad variety of contexts, ranging from gas transport \cite{CroninPritchardRMP09} to the scattering of subatomic particles by atoms or molecules~\cite{RauchQuantumPhen95, BorghesaniIEEE06} to cosmic ray showers~\cite{AnchordoquiAnPhys04}. Inherent to the concept of multiple matter wave scattering is the notion of a refractive index, $n$, which, by the optical theorem~\cite{LLIII}, is proportional to the forward scattering amplitude, $f(0)$. While Im$[f(0)]$ and thus Im$[n]$ are related to the total integral scattering cross section and so can be measured in a standard beam scattering experiment, Re$[f(0)]$ and thus Re$[n]$ reflect the change of the wave's phase velocity and can only be accessed by matter wave interferometry. The first such interferometric measurement was carried out in 1995 by  Schmiedmayer~\textit{et al.}~\cite{SchmiedmayerBookQuOpt,SchmiedmayerPRL95}, who determined both the attenuation and the phase velocity change of sodium atom waves propagating through a number of gases. 
Subsequently, the Pritchard group was able to observe glory-type oscillations in the dependence of the refractive index on the sodium beam velocity~\cite{SchmiedmayerBookInterferometry, HammondBrazJPhys97, RobertsPRL02}. In 2007, Jacquey~\textit{et al.}~\cite{JacqueyPRL07} implemented an improved, Mach-Zehner type of an atom interferometer and measured the index of refraction of lithium waves passing through noble gases. In 2008, Champenois~\textit{et al.}~\cite{ChampenoisPRA08} and, independently, Hornberger and Vaccini~\cite{HornbergerPRA08}, provided an analysis of how the motion of the scatterers affects the measured refractive index. This work showed that the formulae used to extract the refractive index from the experimental data~\cite{SchmiedmayerPRL95, ViguePRA95} and in the related theoretical treatments~\cite{ForreyPRA96,ChampenoisJdePII97,KharchenkoPRA01,ForreyJPB02} lacked consistency with the Beer-Lambert law and thus were incorrect.

According to the scenario of Ref.~\cite{ChampenoisPRA08}, an atomic or ionic beam of ``projectiles'' $p$ of mass $m_p$  and the laboratory velocity $\mathbf{v}_p$ propagates through a gas of density $N_t$ made out of ``target'' particles $t$ of mass $m_t$ and laboratory velocity $\mathbf{v}_t$. For plane matter waves, the velocities $\mathbf{v}_{p,t}$ are related to the laboratory wave vectors $\mathbf{k}_{p,t}$ by $
	\hbar \mathbf{k}_{p,t} = m_{p,t} \mathbf{v}_{p,t}$.
The relative motion of the target and projectile particles is described by the relative wave vector $\mathbf{k}_r = \mu \mathbf{v}_r/\hbar$, where $\mathbf{v}_r = \mathbf{v}_p - \mathbf{v}_t$ is the relative velocity, and $\mu = m_p m_t / (m_p + m_t)$ the reduced mass. 
Taking this scenario into account, Champenois~\textit{et al.}~\cite{ChampenoisPRA08} derived a new formula for the refractive index of a matter wave propagating through a dilute gas
\begin{equation}
	\label{RefrIndex}
	n = 1 + 2\pi N_t \frac{m_p + m_t}{m_t}  \frac{ \left \langle f(k_r,0) \right \rangle}{k_p^2},
\end{equation}
where $f(k_r,0)  \equiv f(0)$ is the forward scattering amplitude pertaining to the scattering angle $\vartheta =0$. The averaging in Eq. (\ref{RefrIndex}) is carried out over the relative wave vectors, $k_r$, corresponding to a normalized relative velocity distribution~\cite{ChampenoisPRA08}, see below.

Previous theoretical treatments of the refractive index of atomic waves, surveyed in Ref. ~\cite{ChampenoisPRA08}, were based on WKB~\cite{SchmiedmayerPRL95} or eikonal~\cite{RobertsPRL02, BlanchardPRA03} approximations, which become analytic only if the repulsive, short-range interaction, is neglected, cf. also refs.~\cite{SchmiedmayerPRL95,ChampenoisPRA08}. In this contribution, we combine the WKB treatment of scattering for the long-range attraction with the Fraunhofer approximation for the short-range repulsion and thereby obtain an analytic model of the refractive index which accounts for both short- and long-range multiple matter-wave scattering. The resulting analytic refractive index provides an additional insight into multiple scattering and facilitates data analysis. 

After briefly introducing the model, we apply it to the refractive index of Li atom waves propagating through Ar, Kr, and Xe gases and compare the results with the measurements of Jacquey~\textit{et al.}~\cite{JacqueyPRL07}. In addition, we exemplify the model's scope by examining the refraction of a Na$^+$ ionic beam passing through a dilute N$_2$ gas.

\section{Model of matter-wave refraction}

In the model introduced herein, we consider the scattering amplitude to consist of two parts:
\begin{equation}
	\label{ScatAmplSplit}
         f (k_r,0) = f_\text{short}(k_r,0) + f_\text{long} (k_r,0) 
\end{equation}
The short range part, $f_\text{short}(k_r,0)$, arises from the scattering by the ``hard core,'' repulsive branch of the potential, and can be evaluated in closed form within the Fraunhofer model of matter wave scattering. The Fraunhofer model was introduced by Drozdov~\cite{DrozdovJETP55} and generalized by Blair~\cite{BlairChapter} in the late 1950s  to treat inelastic nuclear scattering, and adapted by Faubel~\cite{FaubelJCP84} to account for rotationally inelastic thermal collisions between helium atoms and N$_2$ and CH$_4$ molecules.  Recently, we generalized the Fraunhofer model to treat atom-molecule collisions in electric~\cite{LemFriJCP08,LemFriJPCA09}, magnetic~\cite{LemFriPRA09}, and laser fields~\cite{LemFriIJMS09}, and also to gain insight into the stereodynamics of molecular collisions~\cite{LemFriPCCP10,LemJamMirFriJCP10}. 

Within the Fraunhofer model, the hard core part of the potential amounts to a two-dimensional, impenetrable obstacle with sharp boundaries. The elastic scattering amplitude is given by the amplitude for the Fraunhofer diffraction of matter waves by such an obstacle~\cite{LemFriJCP08},
\begin{equation}
	\label{ElasticAngle}
	f_\text{short}(k_r, \vartheta) = i k_r R_0^2(k_r) \frac{J_1 [k_r R_0(k_r) \vartheta]}{k_r R_0(k_r) \vartheta},	
\end{equation}
where  $J_1$ is the Bessel function~\cite{WatsonBesselBook} and $R_0$ is the radius of the interaction which, for an atom-atom potential, is given by the solution of the equation  $V(R_0) = \hbar^2 k_r^2/(2\mu)$, with $V(r)$ the interaction potential. For forward scattering, the amplitude of Eq.~(\ref{ElasticAngle}) becomes
\begin{equation}
	\label{ElasticAmpl}
	f_\text{short}(k_r,0) = \frac{i k_r R_0^2 (k_r)}{2}.
\end{equation}
Hence the Fraunhofer amplitude is purely imaginary and, therefore, contributes only to the imaginary part of the refractive index.

In order to account for the long-range part of the scattering amplitude, $f_\text{long} (k_r,0)$, we make use of the WKB approximation, which is accurate for thermal collisions dominated by partial waves with large angular momenta $l$. We consider the general case of the inverse-power long-range potential,
\begin{equation}
	\label{LRpotential}
	V_\text{long} (r) = - \frac{C_\beta}{r^\beta},
\end{equation}
with $C_\beta > 0$ and $\beta>3$. For an atom-atom interaction, $\beta=6$, while $\beta=4$ for an ion interacting with an atom or molecule. For the potential of Eq.~(\ref{LRpotential}), the WKB phase shift for the $l$-th partial wave is given by~\cite{LLIII}
\begin{equation}
	\label{WKBPhaseShift}
	\delta_l^{(\beta)} = \frac{\mu C_\beta k_r^{\beta-2}}{2 \hbar^2 l^{\beta-1}} \frac{\Gamma(\tfrac{1}{2}) \Gamma(\tfrac{\beta-1}{2})}{\Gamma(\tfrac{\beta}{2})}.
\end{equation}
The WKB forward scattering amplitude ~\cite{LLIII},
\begin{equation}
	\label{WKBampl}
	f_\text{WKB}^{(\beta)} (k_r,0) = \frac{1}{i k_r} \int_0^\infty l \left( e^{2 i \delta_l^{(\beta)}} - 1 \right) dl
\end{equation}
then becomes, on substituting from Eq. (\ref{WKBPhaseShift}):
\begin{multline}
	\label{ScatAmpl}
	f_\text{long}^{(\beta)} (k_r,0) = k_r^\frac{\beta-3}{\beta-1}  \left( \frac{\sqrt{\pi}}{2} \right)^{\frac{\beta+1}{\beta-1}} \left[ \frac{\mu C_\beta}{\hbar^2} \frac{\Gamma(\frac{\beta-1}{2})}{\Gamma(\frac{\beta}{2})} \right]^\frac{2}{\beta-1} \\
	\times \left[ \frac{\Gamma(\frac{3-\beta}{2-2\beta})}{\Gamma(\frac{1}{1-\beta})} + i \frac{\Gamma(\frac{\beta-2}{\beta-1}) }{\Gamma(\frac{\beta+1}{2\beta-2} ) }  \right]
\end{multline}

\section{Refraction of atom matter waves}
\label{refraction}

In the case of an atom beam passing through an atomic gas, $\beta=6$ and the forward scattering amplitude becomes

\begin{multline}
	\label{ScatAmplN6}
	f_\text{atom}(k_r,0) = f_\text{short}(k_r,0) + f_\text{long}^{(6)}(k_r,0) \\
	 = \frac{i k_r R_0^2 (k_r)}{2} + \left( \frac{i k_r}{2} \right)^{3/5} \Gamma\left(\frac{3}{5}\right) \left( \frac{3 \pi}{16} \frac{\mu C_6}{\hbar^2} \right)^{2/5}
\end{multline}
By substituting from Eq.~(\ref{ScatAmplN6}) into Eq.~(\ref{RefrIndex}) and making use of the distribution function given by Eq. (A3) of Ref.~\cite{ChampenoisPRA08}, 
\begin{equation}
	\label{VelocityDistribution}
	P(v_r) = \frac{2 v_r}{\pi^{1/2} \alpha v_p} \exp \left[ - \frac{v_p^2 + v_r^2}{\alpha^2} \right] \sinh \left[ \frac{2 v_p v_r}{\alpha^2} \right]
\end{equation}
with $\alpha = \sqrt{2 k_B T/m_t}$, $k_B$ Boltzmann's constant, and $T$ the temperature, we obtain the following expression for the reduced index of refraction,
\begin{multline}
	\label{RefrIndexN6}
	\eta_\text{atom} \equiv \frac{n-1}{N_t} \\
	 =  2 \pi \frac{\hbar}{m_p v_p^2} \left[ i \langle v_r \rangle R_0^2  + \langle v_r^{3/5} \rangle \left( \frac{i}{2} \right)^{3/5} \Gamma\left(\frac{3}{5}\right) \left( \frac{3 \pi}{16} \frac{C_6}{\hbar} \right)^{2/5}    \right]
\end{multline}
where the velocity averages are given by
\begin{equation}
	\label{AverVelocity}
	\langle v_r \rangle = \int_0^\infty v_r P(v_r) dv_r,
\end{equation}
\begin{equation}
	\label{AverVelocity35}
	\langle v_r^{3/5} \rangle = \int_0^\infty v_r^{3/5} P(v_r) dv_r.
\end{equation}

\begin{figure}
\includegraphics[width=8cm]{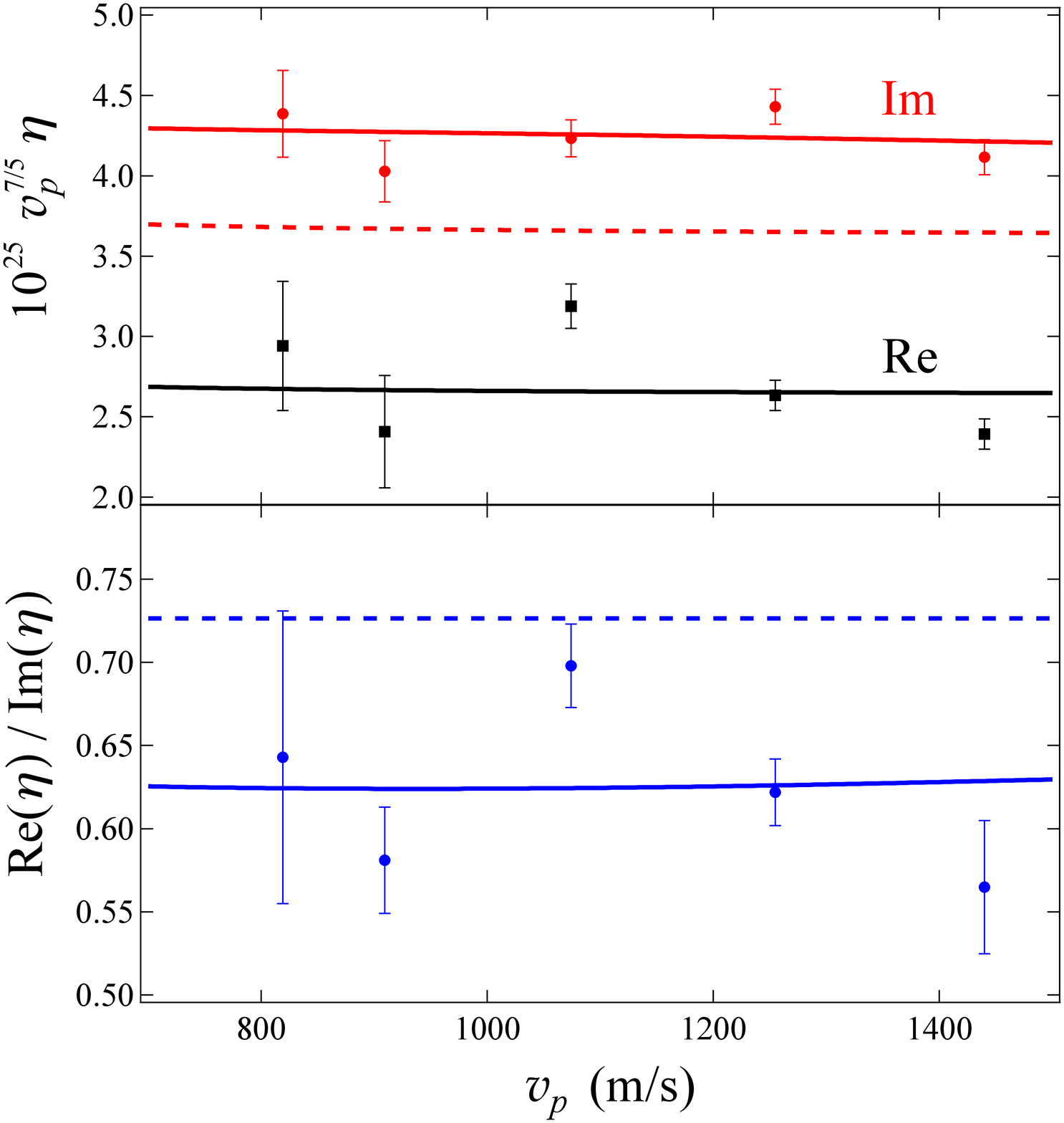}
\caption{\label{fig:LiXeRefrIndex} (Color online) Upper panel shows the real and imaginary parts of the reduced refractive index $\eta \equiv (n-1)/N_t$ [m$^3$] multiplied by $10^{25} v_p^{7/5}$ for Li beam propagating through Xe gas, in dependence of the beam velocity $v_p$. Lower panel shows the corresponding ratio $\text{Re}~(\eta) / \text{Im}~(\eta)$. Model results (solid curves) are compared with experimental data of Ref.~\cite{JacqueyPRL07}. Dashed curves show the results of the model with the short-range potential excluded.}
\end{figure}

\begin{table}
\centering
\caption{Reduced index of refraction, $\eta \equiv (n-1)/N_t$ [m$^3$], for a Li beam propagating through different noble gases with a mean velocity $v_p=1075$~m/s.}
\vspace{0.2cm}
\label{table:LiRefrIndex}
\begin{tabular}{| c | c | c | c | }
\hline 
& Experiment & Model & Model  \\
& of Ref.~\cite{JacqueyPRL07} &   & $f_\text{short}(k_r,0) \equiv 0$ \\[5pt]
\hline
\multicolumn{4}{| c | }{Ar}  \\
\hline & & & \\
$10^{29} \text{Re}(\eta)$  & $1.20\pm0.11$  & 0.97 & 0.97 \\[5pt]
$10^{29} \text{Im}(\eta)$  & $2.11\pm0.06$ & 1.65  & 1.33  \\[5pt]
$\text{Re}(\eta)/\text{Im}(\eta)$  & $0.56\pm0.05$ & 0.58 & 0.73  \\[5pt]
\hline
\multicolumn{4}{| c | }{Kr}  \\
\hline & & & \\
$10^{29} \text{Re}(\eta)$  & $1.57\pm0.10$  & 1.25 & 1.25 \\[5pt]
$10^{29} \text{Im}(\eta)$  & $1.99\pm0.07$ & 2.04  & 1.71  \\[5pt]
$\text{Re}(\eta)/\text{Im}(\eta)$  & $0.78\pm0.04$ & 0.61 & 0.73  \\[5pt]
\hline
\multicolumn{4}{| c | }{Xe}  \\
\hline & & & \\
$10^{29} \text{Re}(\eta)$  & $1.82\pm0.07$  & 1.52 & 1.52 \\[5pt]
$10^{29} \text{Im}(\eta)$  & $2.40\pm0.06$ & 2.42  & 2.08  \\[5pt]
$\text{Re}(\eta)/\text{Im}(\eta)$  & $0.70\pm0.03$ & 0.62 & 0.72  \\[5pt]
\hline
\end{tabular}
\end{table}

In general, the $R_0$ value depends on the relative velocity $v_r$, but for narrow velocity distributions, such as those implemented in the experiments of Jacquey~\textit{et al.}~\cite{JacqueyPRL07}, the dependence is found to be negligible. We could thus simplify the resulting expression by fixing $R_0 (v_r)$ to a constant value pertaining to the mean relative velocity, $R_0 \equiv R_0 (\langle v_r \rangle)$.

While Table \ref{table:LiRefrIndex} compares the experimental values of $\eta_{\text {atom}}$ obtained by Jacquey {\it et al.}~\cite{JacqueyPRL07} with our model results for Li matter waves propagating through Ar, Kr, and Xe gases, Figure 1 singles out the Li+Xe system and compares the experimental and theoretical dependence of the refractive index on the Li velocity. As an input for the analytic model we used potential energy curves of Ahokas~\textit{et al.}~\cite{AhokasJCP2000}. The analytic refractive index is seen to be in a compelling agreement with the experiment. We note that the interference of the matter waves scattered by the
short- and long-range parts of the interatomic potential causes the
refractive index to exhibit glory-type oscillations. These are not correctly 
rendered by our analytic model, as the approximate scattering amplitudes, which were obtained by different methods, fail to ``keep track'' of one another. 

Also presented in both Table \ref{table:LiRefrIndex} and Figure 1 are the values of the refractive index for the long range potential included but the short-range interaction excluded, i.e., for $ f(k_r,0) =f_\text{long}(k_r,0)$, cf. Eq.~(\ref{ScatAmplSplit}). One can see that in such a case the imaginary part of the refractive index is substantially less than in the experiment.

In the limit of a cold atomic gas, i.e., for $T \to 0$, the velocity distribution~(\ref{VelocityDistribution}) becomes a $\delta$-function, which simplifies Eq.~(\ref{RefrIndexN6}), since then $\langle v_r \rangle \to v_p$ and $\langle v_r^{3/5} \rangle \to v_p^{3/5}$. In this case, Re[$\eta_\text{atom}$] is proportional to $v_p^{-7/5}$, while Im[$\eta_\text{atom}$] is proportional to both $v_p^{-1}$ and $v_p^{-7/5}$. However, since the $v_p^{-1}$ term arises from the diffraction of matter waves from the hard core of the potential, it dominates $\text{Im}~[\eta_\text{atom}]$ at large collision energies at the expense of the $v_p^{-7/5}$ term, which arises from the long-range attraction.

\section{Refraction of ion matter waves}

For an ion beam passing through an atomic or molecular gas, $\beta=4$ and the scattering amplitude becomes
\begin{multline}
	\label{ScatAmplN4}
	f_\text{ion}(k_r,0) = f_\text{short}(k_r,0) + f_\text{long}^{(4)}(k_r,0) \\
	= \frac{i k_r R_0^2(k_r)}{2} + \left( \frac{i k_r}{2} \right)^{1/3} \Gamma \left(\frac{1}{3}\right) \left( \frac{\pi}{4} \frac{\mu C_4}{\hbar^2} \right)^{2/3}
\end{multline}
leading to a reduced refractive index for an ionic beam:
\begin{multline}
	\label{RefrIndexN4}
	\eta_\text{ion} \equiv \frac{n-1}{N_t} \\ 
	= 2 \pi \frac{\hbar}{m_p v_p^2} \left[ i \langle v_r \rangle R_0^2  + \langle v_r^{1/3} \rangle \left( \frac{i}{2} \right)^{1/3} \Gamma\left(\frac{1}{3}\right) \left( \frac{\pi}{4} \frac{C_4}{\hbar} \right)^{2/3}    \right],
\end{multline}
with
\begin{equation}
	\label{AverVelocity13}
	\langle v_r^{1/3} \rangle = \int_0^\infty v_r^{1/3} P(v_r) dv_r
\end{equation}

We note that in the case of an anisotropic  interaction of, say, an ion or atom with a linear molecule given by a potential $V(r, \theta)$, the $R_0$ value is given by the spherical part of the hard core of the potential~\cite{LemFriJCP08}. In order to extract $R_0$, we first solve the equation $V(R(\theta), \theta) = \hbar^2 k_r^2/(2\mu)$ to obtain the hard core shape $R(\theta)$, which we then expand in a Legendre series,
\begin{equation}
	\label{LegendreExpansion}
	R(\theta) = R_0 P_0 (\cos \theta) + R_1 P_1 (\cos \theta) + R_2 P_2 (\cos \theta)+ \dots,
\end{equation}
from which we deduce the $R_0$ value.

The values of the refractive index for an ion beam passing through a molecular gas, as exemplified by the Na$^+$--N$_2$ system, are listed in Table \ref{table:NaRefrIndex} at an ion projectile velocity $v_p=5000$~m/s. The requisite value of $R_0$ was extracted from the potential energy surface of Ref. ~\cite{SoldanJCP99}. We see that the refractive index is less affected by switching off the repulsive part of the potential for the ion--molecule system than for the atom--atom system, cf. section \ref{refraction}. This is due to the greater strength of the $-C_4r^{-4}$ interaction compared with that of the $-C_6r^{-6}$ potential and the more dominant role it thus plays in determining $n$.

We note that in the limit of zero temperature, $v_r \to v_p$, the real part of the refractive index~(\ref{RefrIndexN4}) becomes proportional to $v_p^{-5/3}$. The imaginary part of the refractive index is determined by two terms, one of which is proportional to $v_p^{-1}$ and the other to $v_p^{-5/3}$.

\begin{table}
\centering
\caption{Reduced index of refraction, $\eta \equiv (n-1)/N_t$ [m$^3$], for a Na$^+$ beam propagating through N$_2$ gas with a mean velocity $v_p=5000$~m/s.}
\vspace{0.2cm}
\label{table:NaRefrIndex}
\begin{tabular}{| c | c | c | }
\hline 
&Model & Model \\ 
& & $f_\text{short}(k_r,0) \equiv 0$ \\[5pt]
\hline & & \\
$10^{29} \text{Re}(\eta)$  &   0.25 & 0.25 \\[5pt]
$10^{29} \text{Im}(\eta)$  &  0.16  & 0.14  \\[5pt]
$\text{Re}(\eta)/\text{Im}(\eta)$  & 1.60 & 1.73  \\[5pt]
\hline
\end{tabular}
\end{table}

\section{Conclusions}

We presented an analytic model of the refraction of atom or ion matter waves passing through atomic or molecular gases. The values of the refractive index furnished by the model were found to be in good agreement with experiments of Jacquey \textit{et al.}~\cite{JacqueyPRL07}. Our analysis has shown that in order to appraise the imaginary part of the refractive index correctly, we need to account for the diffraction by the repulsive hard core part of the interaction potential. This can be achieved by invoking the Fraunhofer model.

Within the Fraunhofer model of matter wave scattering, the elastic scattering amplitude depends solely on the spherical radius, $R_0$, of the scatterer. As a result, an external field (e.g., magnetic), does not affect the elastic cross section furnished by the model. The corollary is that the field would not affect the index of refraction obtained from the analytic model presented herein either. 
\\

\section{Acknowledgements}
We thank Jacque Vigu\'{e} for discussions and to Gerard Meijer for encouragement and support.

\bibliography{References_thesis}
\end{document}